\documentstyle[12pt,epsfig]{article}
\font\tenrm=cmr10

 1
\font\elevenrm=cmr10 scaled\magstep 1

\renewenvironment{thebibliography}[1]
 { \elevenrm
   \begin{list}{\arabic{enumi}.}
    {\usecounter{enumi}     \setlength{\parsep}{0pt}
     \setlength{\itemsep}{3pt} \settowidth{\labelwidth}{#1.}
     \sloppy
    }}{\end{list}}

\begin{document}
\title{ON THE "FIELD-THEORETICAL APPROACH" TO THE NEUTRON-ANTINEUTRON OSCILLATIONS
 IN NUCLEI }
\author{Vladimir ~Kopeliovich$$\footnote{{\bf e-mail}: kopelio@inr.ru},
\\
$$ \small{\em Institute for Nuclear Research of RAS} \\
\small{\em 60-th October Anniversary Prospect 7a, Moscow 117312}}
\maketitle
{\rightskip=2pc
 \leftskip=2pc
 \noindent}
{\rightskip=2pc
 \leftskip=2pc
\tenrm\baselineskip=11pt
\begin{abstract}
It is argued that within the correct treatment of analytical
properties of the transition amplitudes, in particular, the second order pole structure,
characteristic for the $n - \bar n$ transition in nuclei,
the "infrared divergences" discussed in some papers, do not appear.
Explicit calculation with the help of diagram technique shows
that the neutron-antineutron oscillations are strongly suppressed within a deuteron,
as well as within arbitrary nucleus,
in comparison with the oscillations in vacuum. 
\end{abstract}
 \noindent
\vglue 0.3cm}
\large
{\bf 1.} The neutron-antineutron transition induced by the baryon number violating interaction
$(\Delta B=2)$ predicted within some variants of grand unified theories (GUT) has been discussed
in many papers since 1970 \cite{kuz}, see \cite{mm,ckk,san,dgr,alb}..
Experimental results of searches for such transition are available, in vacuum (reactor experiments
\cite{bal}, and references therein), in nucleus $^{16}O$ \cite{tak} and in $Fe$ nucleus \cite{berg}, 
see also the PDG tables.

During the later time there have been many speculations
that the neutron-antineutron oscillations in nuclei are not
suppressed in comparison with the $n -\bar n$ transition in vacuum \cite{vin1,vin2}.
The arguments were based on the "true field-theoretical approach" to this problem.
The result of \cite{vin1} has been criticized in a number of papers \cite{aga94,mik,lak,aga} which used somewhat different
approaches (potential, S-matrix, diagram), and general physics arguments. 

However, in view of continuing publications \cite{vin2} containing same statement as in \cite{vin1}, it
seems to be necessary to analyze this problem just within the quantum field theory based approach used in \cite{vin1,vin2}.
My consideration is close to the approach of paper \cite{lak} where the diagram technique has been applied
to study neutron-antineutron transition in nuclei, although differs from \cite{lak} in some details.
More recent realistic calculations of the neutron-antineutron tansition in nuclei can be found in \cite{bzk,
faga}.

Here we give first some general arguments based on analytical properties of amplitudes in
favour of suppression of $n-\bar n$ transition in nuclei (section 3). The simplest example of the deuteron 
where the final result  
can be obtained in closed form, is considered in details in section 4, and the result of \cite{lak} 
for the case of the deuteron is reproduced. The analogy with the nucleus formfactor at zero momentum transfer
is noted in section 5. \\

{\bf 2.} To introduce notations, let us consider first the $n\bar n$ transition in vacuum
which is described by the baryon number violating interaction (see, e.g. \cite{mm,mik,lak}) 
$V= \mu_{n\bar n} \sigma_1/2$, $\sigma_1$ being
the Pauli matrix. $\mu_{n\bar n}$ is the parameter which has the dimension of mass, to be predicted
by grand unified theories and to be defined experimentally \footnote{There is relation $\mu_{n\bar n}=2\delta m$
with the parameter $\delta m$ introduced in \cite{mm}. 
The neutron-antineutron oscillation time in vacuum is $\tau_{n\bar n}=1/\delta m=2/\mu_{n\bar n}$, see
also \cite{faga} and references in this paper.}.
The $n-\bar n$ state is described by 2-component spinor $\Psi$,
lower component being the starting neutron, the upper one - the appearing antineutron.
The evolution equation is
$$ i{d\Psi \over dt} = (V_0+V) \Psi \eqno (1) $$
with $V_0=m_N-i\gamma_n/2$ in the rest frame of the neutron ($m_N$ is the nucleon mass, $\gamma_n$ -
the (anti)neutron normal weak interaction decay width, and we take $\gamma_{\bar n}=\gamma_n$, as it follows from
$CP$-invariance of weak interactions). Eq. $(1)$ has solution
$$ \Psi (t) = exp \left[-i\left(\mu_{n\bar n} t\,\sigma_1/2+V_0t\right)\right] \Psi_0 =
\left[cos{\mu_{n\bar n} t\over 2} -i \sigma_1 sin{\mu_{n\bar n} t\over 2}\right]exp(-iV_0t) \Psi_0, \eqno (2) $$
Here $\Psi_0$ is the starting wave function, e.g. $\Psi_0=(0,1)^T$.
In this case we have for an arbitrary time
$$\Psi (\bar n, t) = -i\,  sin{\mu_{n\bar n} t\over 2}exp(-iV_0t), \quad \Psi(n,t) = 
cos{\mu_{n\bar n} t\over 2} exp(-iV_0t) ,\eqno (3) $$
which describes oscillation $n-\bar n$.
Evidently, for large enough observation times, $t^{obs}\gg 1/\mu_{n\bar n}$, the average probabilities
to observe neutron and antineutron are equal if we neglect the natural decay of the neutron (antineutron):
$$ \overline W(\bar n)=\overline{|\Psi(\bar n)|^2}  = \overline{|\Psi (n)|^2} = \overline W(n)=1/2 . \eqno (4)$$
This case is, however, of academic interest, only, since $\gamma_n \gg \mu_{n\bar n}$
\footnote{It is a matter of simple algebra to calculate the integrals over time of the probabilities
$|\Psi(n,t)|^2$ and $|\Psi(\bar n,t)|^2$:.
$$ \int_0^\infty |\Psi(n,t)|^2 dt = {2\gamma_n^2 + \mu_{n\bar n}^2 \over 2\gamma_n(\gamma_n^2 + \mu_{n\bar n}^2)} ,
\quad \int_0^\infty |\Psi(\bar n,t)|^2 dt = {\mu_{n\bar n}^2 \over 2\gamma_n(\gamma_n^2 + \mu_{n\bar n}^2)} $$
for neutron as initial state and for arbitrary, but different from zero $\gamma_n$. The difference between
both quantities is obvious, and disappears when $\gamma_n \to 0$.}
It should be stressed that in the vacuum neutron goes over into antineutron, also discrete 
localized in space state, which can go over again to the neutron, so the oscillation
neutron to antineutron and back takes place.

Since the parameter $\mu_{n\bar n} $ is small, expansion of $sin$ and $cos$ can be made in Eq. $(3)$ at
not too large times. In this case the average (over the time $t^{obs} \ll 1/\mu_{n\bar n}$) change of the 
probability
of appearance of antineutron in vacuum is (for the sake of brevity we do not take into account the (anti)neutron
natural instability which has obvious consequences)
$$ W(\bar n;t^{obs})/t^{obs}= |\Psi (\bar n, t^{obs})|^2/t^{obs} \simeq {\mu_{n\bar n}^2t^{obs}\over 4} \eqno (5) $$
which has, obviously, dimension of the width $\Gamma$.
So, in vacuum the transition $n\to \bar n$ is suppressed if the 
observation time is small, $t^{obs} \ll 1/\mu_{n\bar n}$.
From existing data obtained with free neutrons from reactor the oscillation time is greater than $0.86^.10^8 sec 
\simeq 2.7$ years \cite{bal}, therefore,
$$ \mu_{n\bar n} <  1.5^. 10^{-23} \; eV, \eqno (6) $$
very small quantity.

Recalculation of the quantity  $ \mu_{n\bar n}$ or $\tau_{n\bar n}$ from existing data on nuclei stability \cite{tak,berg}
is somewhat model dependent, and different authors obtained somewhat different results, within
about 1 order of magnitude, see
e.g. discussion in \cite{lak,bzk,faga}. Most recent results  for $ \mu_{n\bar n}$
obtained from nuclear stability data are close to $(6)$ \cite{bzk,faga}, see also next section. \\

 {\bf 3.} In the case of nuclei the $n-\bar n$ line with the transition amplitude $\mu_{n\bar n}$ 
is the element of any amplitude describing the nucleus decay $A \to (A-2) +\, mesons$, where
$(A-2)$ denotes a nucleus or some system of baryons with baryonic number $A-2$, see Fig. 1.                                                                 
The decay probability is therefore proportional to $\mu_{n\bar n}^2$,
and we can write by dimension arguments
$$ \Gamma (A \to (A-2)\, +\, mesons) \sim {\mu_{n\bar n}^2 \over m_0},\eqno (7) $$
where $m_0$ is some energy (mass) scale.
For the result of \cite{vin1,vin2} to be correct the mass $m_0$ should be very small,
$m_0 \sim \mu_{n\bar n} \sim 10^{-23}\, eV$, but we shall argue that $m_0$ is of the order of normal hadronic 
or nuclear scale, $m_0 \sim m_{hadr}\sim (10  - 100)\,MeV$.
We can obtain the same result from the above vacuum formula $(5)$, if we take the time $t^{obs} \sim 1/m_{hadr}$.

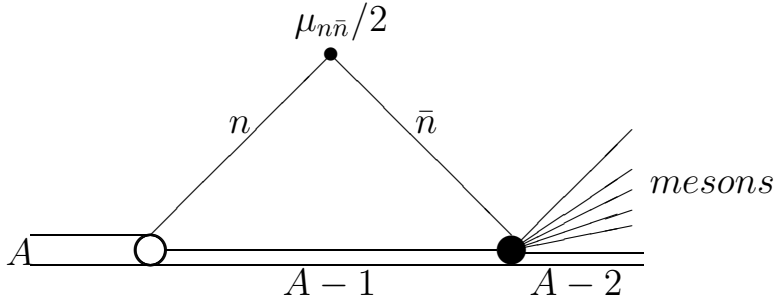
\begin{figure}[h]
\label{aa}
\setlength{\unitlength}{.8cm}
\begin{flushleft}
\begin{picture}(12,6)
\put(3,2){\line(1,0){8.}}

\put(3,2.5){\line(1,0){2.}}
\put(5.25,2.25){\line(1,0){5.5}}

\put(5,2.25){\circle{0.5}}

\put(5,2.25){\circle{0.51}}
\put(5,2.25){\circle{0.517}}
\put(5,2.25){\circle{0.49}}
\put(5,2.25){\circle{0.48}}
\put(11.,2.25){\circle*{0.5}}

\put(11.,2.25){\line(2,1){2.}}
\put(11.,2.25){\line(1,1){2.}}
\put(11.,2.25){\line(5,1){2.}}
\put(11.,2.25){\line(3,2){2.}}
\put(11.,2.25){\line(3,1){2.}}
\put(11.,2.){\line(3,0){2.2}}
\put(11.2,2.2){\line(3,0){2.}}

\put(8,5.5){\circle*{0.2}}
\put(7.4,5.9){$\mu_{n\bar n}/2$}

\put(2.6,2){$A$}
\put(7.2,1.5){$A-1$}

\put(13.3,3.2){$mesons$}
\put(11.3,1.5){$A-2$}
\put(6.3,4.2){$n$}
\put(9.4,4.2){$\bar n$}

\put(5.,2.5){\line(1,1){3}}
\put(8.,5.5){\line(1,-1){3}}

\end{picture}
\vglue 0.1cm
\caption{The diagram describing $n - \bar n$ oscillation in nucleus $A$ with
subsequent annihilation of antineutron to mesons. The final state has the baryon number $A-2$.}
\end{flushleft}
\end{figure}

Indeed, the matrix element of any diagram containing such transition
$$ T(A \to (A-2)\, +\, mesons) \sim$$ $$\sim \mu_{n\bar n}(A-Z)\int V(A;n,(A-1)){\tilde T (\bar n + (A-1) \to (A-2)\, +\, mesons)
\over (E_n-E_n^0+i\delta)^2} dE_n \simeq $$ 
$$\simeq -2\pi i (A-Z) {d(V \,\tilde T)\over dE_n}{(E_n=E_n^0)},  \eqno (8) $$
according to the Cauchy theorem known from the theory of functions of complex variable.
 $E_n$ is the neutron (antineutron) energy - integration variable, $E_n^0$ is the (anti)neutron 
on-mass-shell energy $E_n^0 \simeq m_N + \vec p^2/2m_N$. The
energy-momentum conservation should be taken into account for the vertex $V(A \to n +(A-1))$ which
includes the propagator of the $(A-1)$ system, and for the annihilation amplitude $\tilde T$.
The case of the deuteron considered below is quite transparent and illustrative.

The amplitude $\tilde T$ which describes the annihilation 
of the antineutron, and the vertex function $V$ are of normal hadronic or nuclear scale and cannot, in principle, contain a very small factors
in denominator (or very large factors, of the order of $10^{15}$, in the numerator).
By this reason we come to the above Eq. $(7)$, and the resulting decay width of the nucleus is very small, 
$$ \Gamma (A \to (A-2) +\, mesons) <   10^{-30}\mu_{n\bar n},  \eqno (9)$$
at least $30$ orders of magnitude smaller than the inverse time of neutron-antineutron oscillation
in vacuum $\mu_{n\bar n}$. From Eq. $(7)$ or $(9)$ we obtain
$$ \mu_{n\bar n} \sim \sqrt{\Gamma (A \to (A-2)\, +\, mesons) m_0},\eqno (10) $$
and when one tries to get restriction on $ \mu_{n\bar n}$ from data on nuclei stability \cite{tak,berg}
the result is close to that from vacuum experiment \cite{bal}, somewhat smaller, within one order of magnitude
\cite{dgr,alb,lak}. The result of \cite{bzk}  
differs from that of \cite{lak} for heavier nuclei, and the authors \cite{bzk} come to the conclusion,
that experiments with free neutrons from reactor could provide stronger restriction on the
neutron-antineutron transition parameter than experiments on stability of nuclear matter \footnote{There is,
in fact some kind of competition between both methods, and final result will depend on the progress
to be reached in both branches of experiments --- with free neutrons and with neutrons bound in nuclei.
Friedman and Gal \cite{faga} obtained the restriction $\tau_{n\bar n} > 3.3^.10^8 sec$ from the latest datum on $^{16}O$ 
stability and using the potential approach. Experiments with ultracold neutrons in a trap have been proposed and discussed 
in \cite{ckk,kkl}, but not performed till now.}.

According to \cite{vin1,vin2} the probability of the nucleus decay is proportional to 
$W(t^{obs})\sim \mu_{n\bar n}^2\left(t^{obs}\right)^2$ (the process proceeds similar to the vacuum case), 
where $t^{obs}$ is the large observation time, of the order of $\sim 1$ year or greater.
By this reason the extracted value of $\mu_{n\bar n}$ is
smaller than that given by Eq. $(10)$, by about $15$ orders of magnitude.
Technical reason for strange result obtained in \cite{vin1,vin2} is the wrong interpretation
of the second order pole structure of any amplitude containing the $n - \bar n$ transition.
Instead of using the well developed Feynman diagram technique, the author \cite{vin1,vin2} 
tries to construct the space-time
picture of the process by analogy with the vacuum case, which is misleading, see also discussion 
in conclusions.\\

{\bf 4.} We continue our consideration with the case of the deuteron which is quite simple and 
instructive, and can be treated using
standard diagram technique \footnote{It has been considered in fact in \cite{vin} within reasonable framework
of diagram technique, however, the author has 
drawn later wrong conclusions from this consideration.}.
The point is that in this case there is no final state containing antineutron ---
it could be only the $p\bar n$ state, by charge conservation. But this state is 
forbidden by energy conservation, since the deuteron mass is smaller than the sum of
masses of the proton and antineutron.
Therefore, if the $n-\bar n$ transition took place within the deuteron, the final
state could be only some amount of mesons.
\begin{figure}[h]
\label{deut}
\setlength{\unitlength}{.8cm}
\begin{flushleft}
\begin{picture}(10,6)
\put(3,2){\line(1,0){8.}}
\put(3,2.5){\line(1,0){2.}}

\put(5,2.25){\circle{0.5}}
\put(11.,2.25){\circle*{0.5}}

\put(11.,2.25){\line(3,-1){2.}}
\put(11.,2.25){\line(5,-1){2.}}
\put(11.,2.25){\line(5,1){2.}}
\put(11.,2.25){\line(1,0){2.}}
\put(11.,2.25){\line(3,1){2.}}

\put(8,5.5){\circle*{0.2}}
\put(7.4,5.9){$\mu_{n\bar n}/2$}

\put(2.6,2){$d$}
\put(7.8,1.6){$p$}
\put(13.3,2.2){$mesons$}
\put(6.3,4.2){$n$}
\put(9.4,4.2){$\bar n$}

\put(5.,2.5){\line(1,1){3}}
\put(8.,5.5){\line(1,-1){3}}

\end{picture}
\caption{The diagram describing $n - \bar n$ oscillation in the deuteron with
subsequent annihilation of antineutron and proton to mesons.}
\end{flushleft}
\end{figure}
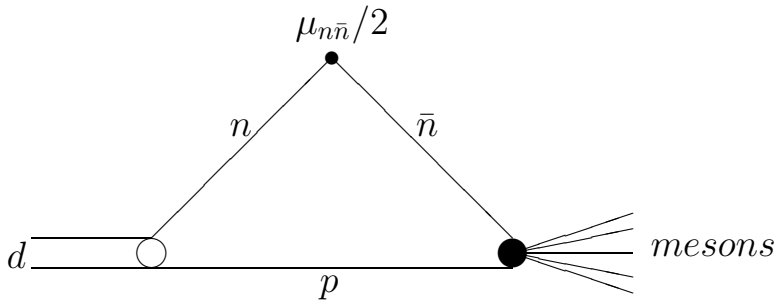
The amplitude of the process is described by the diagrams of the type shown in
Fig. 2 and is equal to
$$ T(d\to mesons) = ig_{dnp}m_N\mu_{n\bar n} \int {T(\bar n p \to mesons)\over (p^2 -m_N^2)[(d-p)^2-m_N^2]^2}
{d^4p\over (2\pi)^4}.  \eqno (11)$$
The constant $g_{dnp}$ is normalized by the condition \cite{ff,cl,ldl}
$${g^2_{dnp}\over 16\pi } = {\kappa\over m_N} = \sqrt{{\epsilon_d\over m_N}}, \eqno (12)$$
which follows, e.g. from the deuteron charge formfactor normalization $F_d(t=0)=1$, see next section.
$\kappa =\sqrt{m_N\epsilon_d}$, $\epsilon_d \simeq 2.22\,MeV$ being the binding energy of the deuteron.
For the vertex $d \to np$ we are writing $2m_Ng_{dnp}$ to ensure the correct dimension of the whole amplitude.

The integration over internal 4-momentum $d^4p$ in $(11)$ can be made easily taking into account the nearest
singularities in the energy $p_0=E$, in the nonrelativistic approximation for nucleons. 
As we shall see right now, the integral over $d^3p$ converges at small $p\sim \kappa$ which corresponds to large distances, $r\sim 1/\kappa$.
By this reason the annihilation amplitude can be taken out of the integration in some average point,
and we obtain the approximate equality
$$ T(d\to mesons) = g_{dnp}m_N\mu_{n\bar n} I_{dNN} T(\bar n p \to mesons) \eqno (11a)$$
with
$$I_{dNN}= {i\over (2\pi)^4}\int {d^4p \over (p^2-m_N^2)[(d-p)^2-m_N^2]^2} \simeq  $$
$$ \simeq {i\over (2\pi)^4 (2m)^3 }\int {d^4p\over (p_0-m_N-\vec p^2/(2m_N) +i\delta)
(m_d -m_N-p_0-\vec p^2/(2m_N) -i\delta)^2} = $$ 
$$=\int {d^3 p \over (2\pi)^3 8m_N[\kappa^2+\vec p^2]^2}={1\over 64\pi m_N\kappa}, \eqno (13) $$
This integral converges at small $|\vec p| \sim \kappa$, more details can be found in the next section.
The decay width (probability) is, by standard technique,
$$ \Gamma (d\to mesons) \simeq \mu_{n\bar n}^2 g_{dnp}^2 I_{dNN}^2 m_N 
\int |T(\bar n p \to mesons)|^2  d\Phi (mesons), \eqno (14) $$
$\Phi (mesons)$ is the final states phase space.
Our final result for the width of the deuteron decay into mesons is
$$\Gamma_{d\to mesons} \simeq {\mu_{n\bar n}^2\over 16\pi\kappa} m_N^2\left[v_0\sigma^{ann}(\bar n p)\right]
_{v_0\to 0}\simeq
{\mu_{n\bar n}^2\over 8\pi\kappa} m_N \left[p_{c.m.}\sigma^{ann}_{\bar n p}\right]_{p_{c.m.}\to 0}, \eqno (15)$$
where $p_{c.m.}$ is the (anti)nucleon momentum in the center of mass system.
This result is very close to that obtained by L.Kondratyuk (Eq. (17) in \cite{lak})
in somewhat different way, using the induced $\bar n p$ wave function \footnote{The result Eq. (17) 
in \cite{lak} can be rewritten in our notations as
$$ \Gamma_{d\to mesons} \simeq 0.01 \mu_{n\bar n}^2 {m_N^2\over \kappa} \left[v_0\sigma^{ann}{\bar n p}\right]
_{v_0\to 0},\qquad \qquad (17')$$
which differs from our result by some numerical factor, close to $1$ and not essential for our conclusions.}.

The annihilation cross section of antineutron with velocity $v_0$ on the proton at rest equals
$$ \sigma (\bar n p \to mesons) = {1\over 4M_N^2v_0}\int |T(\bar n p \to mesons)|^2  d\Phi(mesons).\eqno (16) $$
According to PDG at small $v_0$, roughly, $\left[v_0 \sigma^{ann}_{\bar n p}\right]_{v_0\to 0}\simeq (50 - 55) mb 
\simeq (130 - 140) \,GeV^{-2}$.
So, we obtain $\mu_{n\bar n} \leq 2.5\,10^{-24} eV$, or $\tau_{n\bar n} >\, 
5^.10^8\,sec $ if we take optimistically same restriction for the deuteron stability as it was obtained for
the $Fe$ nucleus, $T_d\simeq T_{Fe} > 6.5^.10^{31} yr$ \cite{berg}.
Our result $(15)$ is valid up to numerical factor of the order $\sim 1$, since we did not consider
explicitly the spin dependence of the annihilation cross section and the spin structure of the incident
nucleus. Same holds in fact for the results obtained in preceeding papers, see \cite{mm,lak} e.g.
                                     
Additional suppression factor in comparison with the case of a free neutron is of the order of
$$\mu_{n\bar n} /\kappa \sim 10^{-31} $$
in agreement with our former rough estimate $(9)$, and disappears, indeed, when the binding 
energy becomes zero \footnote{There is no final formula.
for $\Gamma_{d\to mesons}$ in \cite{vin} to be compared with our result $(14),(15)$. Numerically, however,
the result of \cite{vin} is in rough agreement with our and \cite{lak} estimate.}.
The binding energy of the deuteron should be very small, to give $\kappa \sim \mu_{n\bar n}$, 
to avoid such suppression. At such vanishing binding energy the nucleons inside the deuteron are mostly
outside of the range of nuclear forces, similar to the vacuum case.

Results similar to $(15)$ can be obtained for heavier nuclei, see \cite{dgr,alb,lak,bzk,faga}.
The physical reason of such suppression is quite transparent and has been discussed 
in the literature long ago (see \cite{mm,mik,aga} e.g.):
it is the localization of the neutron inside the nucleus, whereas no localization
takes place in the vacuum case. In the case of the deuteron or heavier nucleus the annihilation of
antineutron takes place, and final state is some continuum state containing mesons.
By this reason the transition of the final state back to the incident nucleus  is not possible
in principle, and there cannot be oscillation of the type, e.g. $d\to mesons \to d$.
This is important difference from the case of the free neutron.\\

{\bf 5.} As we noted previously, the presence of the second order pole in intermediate energy variable 
is characteristic for the processes with
the neutron - antineutron transition, but it is in fact not a new peculiarity, it takes place 
also for the case of 
the nucleus formfactor with zero momentum transfer, $F_A(q=0)$. 
Let us consider as an example the deuteron charge formfactor.   
In the zero range approximation it can be written as
$$ F_d(q) = {i(2m g_{dnp})^2\over (2\pi)^4}\int {d^4p \over (p^2-m_N^2)[(d-p)^2-m_N^2] 
[(d-p+q)^2-m_N^2]}.  \eqno (17)$$
Behind the zero range approximation $g_{dnp}$ should be considered as a function of the relative
$n-p$ momentum, not as a constant.
For $q=0$ second order pole appears, and we come to the expression for $F(q=0)$ containing
the integral $I_{dNN}$ introduced above in Eq. $(13)$:
$$F_d(0) = (2m_Ng_{dnp})^2 I_{dNN}.  \eqno (18) $$

\begin{figure}[h]
\label{form}
\setlength{\unitlength}{.8cm}
\begin{flushleft}
\begin{picture}(10,6.5)
\put(3,1){\line(1,0){8.}}
\put(3,1.5){\line(1,0){2.}}

\put(5,1.25){\circle{0.5}}

\put(11.,1.25){\circle{0.5}}
\put(11.,1.){\line(1,0){2.}}
\put(11.,1.5){\line(1,0){2.}}


\put(8,4.5){\circle{0.3}}
\put(8,4.65){\line(0,1){2.}}
\put(7.4,5.8){$\gamma$}

\put(2.6,1){$d$}
\put(7.8,.6){$n$}
\put(13.,1.){$d$}
\put(6.3,3.2){$p$}
\put(9.4,3.2){$p$}

\put(5.,1.5){\line(1,1){2.85}}
\put(8.106,4.394){\line(1,-1){3}}

\end{picture}
\caption{The diagram describing the deuteron charge formfactor.}
\end{flushleft}
\end{figure}
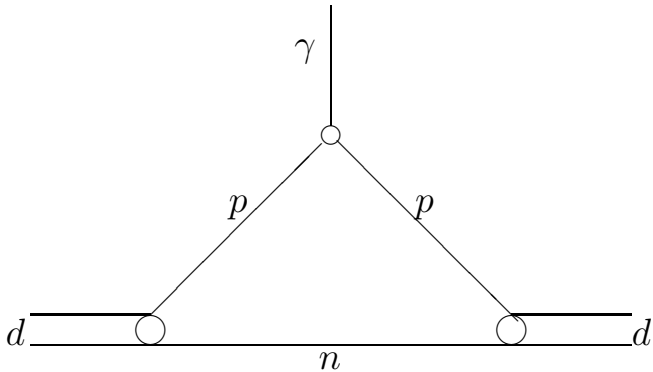

In the nonrelativistic approximation, when only 
the nearest in energy $E=p_0$ singularities are taken into account,
the integral over the energy  has the structure
$$ I_{dNN} \sim \int {dE\over (E-a+i\delta)(E-b-i\delta)^2} = {-2\pi i\over (a-b)^2}, \eqno (19)$$
$a=m_N+\vec p^2/2m_N.\; b=m_d-m_N-\vec p^2/2m_N$, $a-b=\epsilon_d +\vec p^2/m_N$,
and can be calculated using the lower contour, or the upper contour, with the help of formulas
known from the theory of functions of complex variables.
After this we obtain
$$F_d(q=0) = {g_{dnp}^2m_N\over 16\pi^3} \int {d^3p\over (\kappa^2+\vec p^2)^2}. \eqno (20)$$
Since $F_d(0) =1$, this leads to the above mentioned normalization condition $g_{dnp}^2/(16\pi) =\sqrt{\epsilon_d/m_N}$
\footnote{As it is known from nonrelativistic diagram technique, the wave function of the deuteron in
momentum representation is $\Psi_d(\vec p) = g_{dnp}/[4\pi^{3/2} (\kappa^2+\vec p^2)]$, therefore,
the normalization of the charge formfactor $F_d(0)=1$ follows from the normalization of
the deuteron wave function, which is also well known from quantum mechanics.}.

This relation between the constant $g_{dnp}$ and the binding energy of the
weakly bound system (deuteron in our case) is known for a long time \cite{ff,cl,ldl}.
It was obtained in \cite{ff,cl,ldl} using different methods, dispersion relation, for example.

If the infrared divergence discussed in \cite{vin1,vin2} takes place for the process of $n-\bar n$ transition
in nucleus, it should take place also for the nucleus formfactor at zero momentum transfer.
But it is well known not to be the case.
There is no "new limit on neutron - antineutron transition" \cite{vin1};  
instead, one should treat correctly singularities of the transition amplitudes in the
complex energy plane, in particular, the second order pole contribution to the transition
amplitudes. The author of \cite{vin1,vin2} tries to reconstruct the space-time picture of the 
process, but the correspondence of this picture to the well justified amplitude, as it appears
from Feynman diagrams, is questionable.  The infrared divergence discussed in \cite{vin1,vin2}
is an artefact of this inadequate space-time picture of the whole process of $n-\bar n$ transition
with subsequent antineutron annihilation. Another quite unrealistic consequence of this  
space-time picture is the nonexponential law of the nucleus decay.

Field-theoretical description of nuclear reactions and processes is potentially useful, 
it allows to study some effects which is not possible in principle to study in other way, 
e.g. relativistic corrections to different observables. One should be, however, very careful to 
treat adequately analytical  
properties of contributing amplitudes. E.g., in the case of the parity violating amplitude of $np \to d\gamma$ capture
it was necessary to take into account contributions of all singularities (poles) of the amplitude
in the complex energy plane, not only contributions of the nearest poles in energy variable, as it is made
usually in nonrelativistic calculations. The nonrelativistic diagram technique developed up to that time
turned out to be misleading for the case of physics problem considered in \cite{vbk}.
Besides, and it is the spesifics of the processes with
photon emission, the contact terms should be reconstructed to 
ensure the gauge invariance of the whole amplitude of photon radiation \cite{vbk}.\\

{\bf  Acknowledgements.}
Present investigation, performed partly with pedagogical purposes, has been initiated  
by V.M.Lobashev and V.A.Matveev         
who have drawn my attention to the longstanding discrepance between papers \cite{vin1,vin2}
and the results accepted by scientific community (\cite{kuz}-\cite{dgr},\cite{mik,lak,aga} 
and references therein).

I am thankful to A.E.Kudryavtsev for careful reading the manuscript and many valuable remarks. 
I'm indebted also to A.Gal, B.Z.Kopeliovich, M.I.Krivoruchenko, I.K.Potashnikova and to 
participants of the seminars of the INR of RAS for 
useful discussions and comments. Numerous discussions with V.I.Nazaruk did not lead to any change of 
his way of thinking \footnote{In his latest comment \cite{vin10} Nazaruk makes the statement "In the correct model the
neutron line entering the $n\bar n$ transition vertex should be the wave function." It means that new rules
are proposed in \cite{vin1, vin2} instead of well known Feynman rules. These "new rules" should be,
at least, clearly formulated. and, second, these rules should allow to reproduce all well known results of 
nuclear theory, There is no need of further discussion.}.

This work was supported in part by Fondecyt (Chile), 
grant number 1090236.
\\

\end{document}